\documentstyle[amsfonts,preprint,aps]{revtex}

\begin{document}
\title{Teleportation of the U-NOT Gate and Quantum Cloning: a Computational Network
and Experimental Realization}
\author{F. Sciarrino, C. Sias, M. Ricci, and F.\ De Martini}
\address{Dipartimento di Fisica and \\
Istituto Nazionale per la Fisica della Materia\\
Universit\`{a} di Roma ''La Sapienza'', Roma, 00185 - Italy}
\date{\today}
\maketitle

\begin{abstract}
We present a computational circuit which realizes contextually the
Tele-UNOT\ gate and the universal optimal quantum cloning machine (UOQCM).
We report the experimental realization of the probabilistic UOQCM with
polarization encoded qubits. This is achieved by combining on a symmetric
beam-splitter the input qubit with an ancilla in a fully mixed state .
\end{abstract}


Quantum information is encoded in qubits, the quantum analogue of the
classical bits. Transformations of these systems follow the laws of quantum
mechanics. This fundamental requirement implies theoretical limitations as
far as the ''cloning'' and/or ''spin-flipping'' processes of qubits are
concerned. For instance, it has been shown that an arbitrary unknown qubit
cannot be perfectly cloned: $\left| \Psi \right\rangle \rightarrow \left|
\Psi \right\rangle \left| \Psi \right\rangle $, a consequence of the
so-called ``no cloning theorem'' \cite{1}, \cite{2}. Another ''impossible''
device is the quantum NOT gate, the transformation that maps any qubit into
the orthogonal one $\left| \Psi \right\rangle \rightarrow \left| \Psi
^{\perp }\right\rangle $ \cite{3}. In the last years a great deal of
theoretical investigation has been devoted to finding the best
approximations allowed by quantum mechanics to these two processes and to
establish for them the corresponding ''optimal'' values of the ''fidelity'' $%
F<1$.\ This problem has been solved in the general case \cite{4,5,6}. In
particular, it was found that a one-to-two universal optimal quantum cloning
machine (UOQCM), i.e. able to clone one qubit\ into two qubits ($%
1\rightarrow 2$), can be realized with a fidelity $F_{CLON}=\frac{5}{6}$.
The ''flipping'' of the input injected qubit, which realizes a $%
(1\rightarrow 1)$ Universal-NOT gate (U-NOT), can be implemented with a
fidelity $F_{NOT}=\frac{2}{3}$ \cite{5,6}. On the other hand, a peculiar and
fundamental resource available in the field of quantum information is
quantum state teleportation (QST) protocol by which an unknown input qubit $%
\left| \phi \right\rangle _{S}=\alpha \left| 0\right\rangle _{S}+\beta
\left| 1\right\rangle _{S}\ $is destroyed at a sending place (Alice:\ ${\cal %
A}$) while its perfect replica appears at a remote place (Bob:\ ${\cal B}$)
via dual quantum and classical channels \cite{7}. Here we present a linear
scheme that establishes a connection between these three quantum operations.

Previously the phenomena of quantum teleportation, quantum cloning and UNOT
gate have been analyzed and experimentally implemented using different
approaches. The quantum teleportation has been realized using a linear
interaction to couple the qubit $S$ to be teleported with an $EPR$ pair \cite
{8}. Furthermore it has been proposed to implement the optimal quantum
cloning machine by a nonlinear quantum optical ''amplification'' method,
i.e. by associating the cloning effect with the QED\ ''stimulated emission''
process \cite{9}. This proposal was quickly followed by some successful
experimental demonstrations \cite{10,11,12}. It has been argued by \cite{12}
that when the cloning process is realized in a subspace ${\it H}$ of a
larger nonseparable Hilbert space ${\it H}\otimes K$ which is acted upon by
a physical apparatus, the same apparatus performs contextually in the space $%
K$ the ''flipping'' of the input injected qubit, then realizing a U-NOT gate 
\cite{5,6}. As an example, a UOQCM can be realized on one output mode of a
non-degenerate ''quantum-injected'' optical parametric amplifier (QIOPA),
while the U-NOT transformation is realized on the other mode \cite{11,12}.

Very recently a novel and quite unexpected relation between the QST, UOQCM
and U-NOT\ gate processes has been presented \cite{13}. Indeed it has been
shown that it is possible to implement the $1\rightarrow 2$ universal
optimal quantum cloning machine and $1\rightarrow 1$ universal NOT gate by
slightly modifying the protocol of quantum state teleportation. Let us now
briefly outline this protocol. Alice possesses the starting qubit $\left|
\phi \right\rangle _{S}$, and, in analogy to the QST, there is a second
partner, Bob, who shares with her the entangled ''singlet''\ state $\left|
\Psi ^{-}\right\rangle _{AB}=2^{-%
{\frac12}%
}\left( \left| 0\right\rangle _{A}\left| 1\right\rangle _{B}-\left|
1\right\rangle _{A}\left| 0\right\rangle _{B}\right) $. By means of a
projective measurement and of classical communication, Alice obtains two
qubits $S$, $A$ that are the optimal quantum clones of the starting qubit $%
\left| \phi \right\rangle _{S}$, while Bob receives its optimal flipped
state. Let us present in more details the procedure. Alice performs a
dichotomic projective measurement on the systems $S$ and $A$ able to
identify $\left| \Psi ^{-}\right\rangle _{SA}$ and its complementary space.
With a probability $p=\frac{1}{4}$ the state $\left| \Psi ^{-}\right\rangle
_{SA}$ is detected by $A$ and the qubit $\left| \phi \right\rangle $ is
teleported to Bob. With probability $p=\frac{3}{4}$, the qubits $S$ and $A$
are projected in the subspace spanned by $\left\{ \left| \Psi
^{+}\right\rangle _{SA},\left| \Phi ^{-}\right\rangle _{SA},\left| \Phi
^{+}\right\rangle _{SA}\right\} $ which is the symmetric space of $A$ and $%
S. $ The output state is obtained by projecting the initial overall state $%
\left| \Psi ^{-}\right\rangle _{AB}\left| \phi \right\rangle _{S}$\ onto the
subspace orthogonal to $\left| \Psi ^{-}\right\rangle _{SA}\left\langle \Psi
^{-}\right| _{SA}\otimes {\Bbb I}_{B}$, i.e. by applying the projector: 
\begin{equation}
P_{SAB}=({\Bbb I}_{SA}-\left| \Psi ^{-}\right\rangle _{SA}\left\langle \Psi
^{-}\right| _{SA})\otimes {\Bbb I}_{B}  \label{proiettore}
\end{equation}
The reduced density matrices for the qubits $S,A$ and $B$ are then \cite{13}%
: 
\begin{eqnarray}
\rho _{S} &\equiv &Tr_{A}\rho _{SA}=\frac{5}{6}\left| \phi \right\rangle
\left\langle \phi \right| +\frac{1}{6}\left| \phi ^{\perp }\right\rangle
\left\langle \phi ^{\perp }\right| =\rho _{A}\equiv Tr_{S}\rho _{SA} \\
\rho _{B} &\equiv &Tr_{SA}\left| \widetilde{\Omega }\right\rangle
\left\langle \widetilde{\Omega }\right| =\frac{2}{3}\left| \phi ^{\perp
}\right\rangle \left\langle \phi ^{\perp }\right| +\frac{1}{3}\left| \phi
\right\rangle \left\langle \phi \right|
\end{eqnarray}
by which the ''optimal'' values for the fidelities of the two ''forbidden''\
processes $F_{CLON}=\frac{5}{6}$ and $F_{UNOT}=\frac{2}{3}$ are achieved for
the qubits on the Alice and Bob sites respectively\cite{11,12}. Bob
identifies whether the U-NOT gate is realized at his site by reading the
information (1 bit) received by Alice on the classical channel. For example,
such bit can assume the value $1$ if Alice identifies the Bell state $\left|
\Psi ^{-}\right\rangle _{SA}$ and $0$ if not. At the end of the protocol the
universal NOT gate applied to the input qubit has been teleported at Bob's
site and hence this process has been called $Tele-UNOT$ gate. Recently
Gottesman and Chuang have proposed to exploit the teleportation protocol to
transfer quantum gates in order to relax several experimental constraints to
achieve fault-tolerant processing \cite{14}. Indeed the linear optics
quantum computation exploits gate teleportation in order to transform a
probabilistic computation into a near-deterministic one \cite{15}.

The projector $P_{SAB}$ in Eq.(\ref{proiettore}) leads to the realization of
the $UOQCM$ and $Tele-UNOT$ gate, as said. Here we present, in analogy with
the quantum circuit associated with quantum teleportation, a quantum circuit
that achieves this purpose: the projection is obtained combining Hadamard
gates, $C-NOT$ gates, a Toffoli gate, and the projective measurement in the
computational basis.

The box ''EPR preparation'' prepares the singlet state while the ''Quantum
Machine'' box achieves the projection into the symmetric subspace. The
readout of the ancilla qubit $\widetilde{a}$, initially in the state $\left|
0\right\rangle _{\widetilde{a}}$, ensures that the projection into the
symmetric space has been obtained. Let us now analyze in more details the
logic of the quantum circuit. After the EPR preparation, the state of the
overall system is 
\begin{equation}
\left| \phi \right\rangle _{S}\otimes \left| \Psi ^{-}\right\rangle
_{AB}\otimes \left| 0\right\rangle _{\widetilde{a}}=\frac{1}{2}\left[ 
\begin{array}{c}
-\left| \Psi ^{-}\right\rangle _{SA}\left| \phi \right\rangle _{B}-\left|
\Psi ^{+}\right\rangle _{SA}\left( \sigma _{Z}\left| \phi \right\rangle
_{B}\right) \\ 
+\left| \Phi ^{-}\right\rangle _{SA}\left( \sigma _{X}\left| \phi
\right\rangle _{B}\right) +\left| \Phi ^{+}\right\rangle _{SA}\left( \sigma
_{Z}\sigma _{X}\left| \phi \right\rangle _{B}\right)
\end{array}
\right] \otimes \left| 0\right\rangle _{\widetilde{a}}
\end{equation}

The box labelled {\bf (1)} transforms the state $\left| \Psi
^{-}\right\rangle _{SA}$ into $\left| 1\right\rangle _{S}\left|
1\right\rangle _{A},$ while the other three Bell states $\left\{ \left| \Psi
^{+}\right\rangle _{SA},\left| \Phi ^{-}\right\rangle _{SA},\left| \Phi
^{+}\right\rangle _{SA}\right\} $ are respectively transformed into $\left\{
\left| 0\right\rangle _{S}\left| 1\right\rangle _{A},\left| 1\right\rangle
_{S}\left| 0\right\rangle _{A},\left| 0\right\rangle _{S}\left|
0\right\rangle _{A}\right\} .$ By means of a Toffoli gate, the state $\left|
1\right\rangle _{S}\left| 1\right\rangle _{A}$ induces the flipping of the
state of the qubit $\widetilde{a}$ from $\left| 0\right\rangle _{\widetilde{a%
}}$ to $\left| 1\right\rangle _{\widetilde{a}},$ whereas the other states
leave the qubit $\widetilde{a}$ unaltered. Then we get the following state 
\begin{equation}
\frac{1}{2}\left[ 
\begin{array}{c}
-\left| 1\right\rangle _{S}\left| 1\right\rangle _{A}\left| \phi
\right\rangle _{B}\left| 1\right\rangle _{\widetilde{a}}-\left|
0\right\rangle _{S}\left| 1\right\rangle _{A}\left( \sigma _{Z}\left| \phi
\right\rangle _{B}\right) \left| 0\right\rangle _{\widetilde{a}} \\ 
+\left| 1\right\rangle _{S}\left| 0\right\rangle _{A}\left( \sigma
_{X}\left| \phi \right\rangle _{B}\right) \left| 0\right\rangle _{\widetilde{%
a}}+\left| 0\right\rangle _{S}\left| 0\right\rangle _{A}\left( \sigma
_{Z}\sigma _{X}\left| \phi \right\rangle _{B}\right) \left| 0\right\rangle _{%
\widetilde{a}}
\end{array}
\right]
\end{equation}
Finally the action of the box labelled {\bf (2) }restores the initial states
of the qubits $S$ and $A$ leading to: 
\begin{equation}
\left| \Sigma \right\rangle _{SAB\widetilde{a}}=\frac{1}{2}\left[ 
\begin{array}{c}
-\left| \Psi ^{-}\right\rangle _{SA}\left| \phi \right\rangle _{B}\left|
1\right\rangle _{\widetilde{a}}-\left| \Psi ^{+}\right\rangle _{SA}\left(
\sigma _{Z}\left| \phi \right\rangle _{B}\right) \left| 0\right\rangle _{%
\widetilde{a}} \\ 
+\left| \Phi ^{-}\right\rangle _{SA}\left( \sigma _{X}\left| \phi
\right\rangle _{B}\right) \left| 0\right\rangle _{\widetilde{a}}+\left| \Phi
^{+}\right\rangle _{SA}\left( \sigma _{Z}\sigma _{X}\left| \phi
\right\rangle _{B}\right) \left| 0\right\rangle _{\widetilde{a}}
\end{array}
\right]
\end{equation}

If the projective measurement on the ancilla qubit $\widetilde{a}$ gives as
result $1$ the qubits $A$ and $S$ end up in the state $\left| \Psi
^{-}\right\rangle _{SA}$ while the qubit $B$ is in the state $\left| \phi
\right\rangle ,$ which has then been teleported from ${\cal A}$ to ${\cal B}%
. $ On the contrary, if we obtain the result $0$ the overall state becomes 
\begin{equation}
-\left| \Psi ^{+}\right\rangle _{SA}\left( \sigma _{Z}\left| \phi
\right\rangle _{B}\right) +\left| \Phi ^{-}\right\rangle _{SA}\left( \sigma
_{X}\left| \phi \right\rangle _{B}\right) \left| 0\right\rangle _{\widetilde{%
a}}+\left| \Phi ^{+}\right\rangle _{SA}\left( \sigma _{Z}\sigma _{X}\left|
\phi \right\rangle _{B}\right)
\end{equation}
that is equal to\ the state obtained applying the projector $P_{SAB}$ (\ref
{proiettore}) to $\left| \phi \right\rangle _{S}\otimes \left| \Psi
^{-}\right\rangle _{AB}$.\ The result of the ancilla measurement is
communicated to Bob and we realize the optimal quantum cloning machine and
the $Tele-UNOT$ gate of the input qubit $\left| \phi \right\rangle $.

Note that the presence of the entangled state $\left| \Psi ^{-}\right\rangle
_{AB}\ $is not strictly necessary for the implementation of the solely
quantum cloning process. Instead of sharing an EPR pair with Bob, Alice
needs a qubit $A$ prepared in a {\it fully mixed} state $\rho _{A}=\frac{%
{\Bbb I}_{A}}{2}.$ Of course in this case only the Alice's apparatus is
relevant and the U-NOT process is absent. In a previous experiment \cite{13}
we have reported the realization of the $Tele-UNOT$ gate by adopting an
entangled state. Here we want to demonstrate experimentally that the
entanglement is not required for the implementation of quantum cloning.

The qubits $S$ and $A$ have been codified in the polarization states of two
single photons. In this case $P_{SAB}$ can be experimentally achieved with
an Ou-Mandel interferometer $BS_{A}$ \cite{16}. We exploit the Bose mode
coalescence (enhancement) BMC associated to the projection into the
symmetric subspace $\Pi _{sym}.$ Indeed the {\it symmetry} of the projected
subspace by BMC is implied by the intrinsic {\it Bose symmetry} of the 2
photons Fock state realized at the output of $BS_{A}$.

The qubit to be cloned is: $\left| \phi \right\rangle _{S}$ $=\alpha \left|
H\right\rangle _{S}+\beta \left| V\right\rangle _{S}$ where $\left|
H\right\rangle $ and $\left| V\right\rangle $ respectively correspond to the
horizontal and vertical linear polarizations of a single photon injected in
one input mode $S$ of a $50:50\,$beamsplitter $BS_{A}$. A fully mixed state $%
\rho _{A}$ is simultaneously injected on the other input mode $A$ of $BS_{A}$
where the two input modes are linearly superimposed. Consider the overall
output state which is realized on the two output modes $1\,$and $2$ of $%
BS_{A}$. It can be expressed as a linear combination of the Bell states $%
\left\{ \left| \Psi ^{-}\right\rangle _{SA},\left| \Psi ^{+}\right\rangle
_{SA},\left| \Phi ^{-}\right\rangle _{SA},\left| \Phi ^{+}\right\rangle
_{SA}\right\} $. As it is well known, the realization of the singlet $\left|
\Psi _{SA}^{-}\right\rangle $ is unambiguously identified by the detection
of a single photon on each one of the two output modes of $BS_{A}$ while the
realization of the set of the other three Bell states corresponds to the
emission of photon pairs on either one of the output modes \cite{16}. Hence
the detection of two photons over either the mode $1$ or $2$, a
Bose-mode-occupation enhancement, implies the projection by $P_{SA}=\left( 
{\Bbb I}_{SA}-\left| \Psi _{SA}^{-}\right\rangle \left\langle \Psi
_{SA}^{-}\right| \right) $ of the input state into the space orthogonal to $%
\left| \Psi _{SA}^{-}\right\rangle $.

In the present experiment, a pair of non entangled photons with wavelength $%
\lambda =532nm$ and with a coherence-time $\tau _{coh}=80fs$, were generated
by a spontaneous parametric down conversion (SPDC) process in a Type I BBO\
crystal in the initial polarization product state $\left| H\right\rangle
_{S}\left| H\right\rangle _{A}$. The two photons were then injected on the
two input modes $S$ and $A$ of $BS_{A}$. The input qubit $\left| \phi
\right\rangle _{S}$, associated with mode $S$ was polarization encoded by
means of a waveplate (wp) $WP_{S}$. The transformation used to map the state 
$\left| H\right\rangle _{A}$ into $\rho _{A}=\frac{{\Bbb I}_{A}}{2}$ was
achieved by stochastically rotating a $\lambda /2$ waveplate inserted on the
mode $A$ during the experiment. In this way the statistical evolution of $%
\left| H\right\rangle _{A}$ into two orthogonal states with equal
probability was achieved. The photons $S$ and $A$ were injected in the two
input arms of $BS_{A}$ with a mutual delay $\Delta t$ micrometrically
adjustable by a translation stage with position settings $Z=2\Delta tc$. The
setting value $Z=0$ was assumed to correspond to the full overlapping of the
photon pulses injected into $BS_{A}$, i.e. to the maximum photon
interference.

For the sake of simplicity, we only analyzed the measurements performed on
the $BS_{A}$ output mode $1$:\ Fig.2. The polarization state on this mode
was analyzed by the combination of the wp $WP_{C}$ and of the polarization
beam splitter $PBS_{C}$. For each input polarization state $\left| \phi
\right\rangle _{S}$, $WP_{C}$ was set in order to make $PBS_{C}$ to transmit 
$\left| \phi \right\rangle _{S}$ and reflect $\left| \phi ^{\perp
}\right\rangle _{S}$. The ''cloned'' state $\left| \phi \phi \right\rangle
_{S}$ could be detected on mode $1$ by a two-photon counter, realized in our
case by first separating the two photons by a $50:50$ beam splitter $BS_{C}$
and then detecting the coincidence $[D_{A1},D_{A2}]$ between the output
detectors $D_{A1}$ and $D_{A2}$:\ Fig.2. Any coincidence between detectors $%
D_{B}$ and $D_{A2}$ corresponded to the realization of the state $\left|
\phi \phi ^{\perp }\right\rangle _{S}$. First consider the cloning machine
switched off, by setting: $\Delta t>\tau _{coh}$ i.e. by making $S$ and $A$
not interfering on $BS_{A}$. In this case, since the states $\left| \phi
\phi \right\rangle _{S}$ and $\left| \phi \phi ^{\perp }\right\rangle _{S}$
were realized with the same probability on mode $1$, the rate of
coincidences detected by $[D_{A1},D_{A2}]$ and $[D_{A2},D_{B}]\ $were
expected to be equal. By turning on the cloning machine, i.e. by setting $%
\Delta t<<\tau _{coh}$, on mode $1$ the output density matrix $\rho _{SA}$
was realized implying an enhancement by a factor $R=2$ of the counting rate
by $[D_{A1},D_{A2}]$ and no rate enhancement by $[D_{A2},D_{B}]$. The
measurement of $R$ was carried out by coincidence measurements involving
simultaneously $[D_{A1},D_{A2}]$ and $[D_{A2},D_{B}]$. The experimental data
are reported\ in Fig. 3 for three different input states $\left| \phi
\right\rangle _{S}=\left| H\right\rangle $, $2^{-%
{\frac12}%
}(\left| H\right\rangle +\left| V\right\rangle )$, $2^{-%
{\frac12}%
}(\left| H\right\rangle +i\left| V\right\rangle )$. There circle and square
markers refer respectively to the $[D_{A1},D_{A2}]$ and $[D_{A2},D_{B}]$
coincidences versus the position setting $Z.$ We may check that the cloning
process only affects the $\left| \phi \phi \right\rangle _{S}$ component, as
expected and $R\;$is determined as the ratio between the peak values
(cloning machine switched on) and the basis values (cloning machine switched
off). The corresponding experimental values of the {\it cloning fidelity }$F=%
\frac{2R+1}{2R+2}$ are: $F_{H}=0.827\pm 0.002$;$\ F_{H+V}=0.825\pm 0.002$; $%
F_{H+iV}=0.826\pm 0.002$. These ones are in good agreement with the optimal
value $F_{th}=5/6\approx 0.833$ which corresponds to the limit value of the
amplification ratio $R=2.$ We note that, while our experiment did not adopt
an ancilla qubit, the measurement of the number of photons over the output
modes allowed us to identify whether the projection on the symmetric space $%
\Pi _{sym}$ had been obtained.

In conclusion we presented how it is possible to implement the $1\rightarrow
2$ universal optimal quantum cloning machine and $1\rightarrow 1$
Tele-Universal NOT gate by means of a quantum network that can be
implemented for all kind of qubits. We report the highest experimental
fidelity attained so far for the $1\rightarrow 2$ UOQCM adopting single
photon polarization encoded qubits. Furthermore the simplicity of our setup
renders this eavesdropping technique practically available for optimal
attack to cryptographic communications \cite{17}.

This work has been supported by the FET European Network on Quantum
Information and Communication (Contract IST-2000-29681: ATESIT), by Istituto
Nazionale per la Fisica della Materia (PRA\ ''CLON'')\ and by Ministero
dell'Istruzione, dell'Universit\`{a} e della Ricerca (COFIN 2002).

\centerline{\bf Figure Captions}

\vskip 8mm

\parindent=0pt

\parskip=3mm

Figure.1. Realization of $UOQCM$ and $Tele-UNOT$ gate by means of quantum
circuits. The preparation box contains the circuit that generates the state $%
\left| \Psi ^{-}\right\rangle _{AB}$ while the ''quantum machine'' box
represents the apparatus performing the projective measurement over the
antisymmetric and symmetric $\left( \Pi _{sym}\right) $ subspaces of the two
qubits $A$ and $S$. The Toffoli gate can be realized by means of single
qubit gates and $C-NOT$ gates.

Figure.2. Experimental setup for the optical implementation of the UOQCM by
a modified Teleportation protocol.

Figure3. Experimental result of the universal cloning process for different
input qubits corresponding to the encoded polarizations: $\left|
H\right\rangle $, $2^{-1/2}\left( \left| H\right\rangle +\left|
V\right\rangle \right) $ and $2^{-1/2}\left( \left| H\right\rangle +i\left|
V\right\rangle \right) $.

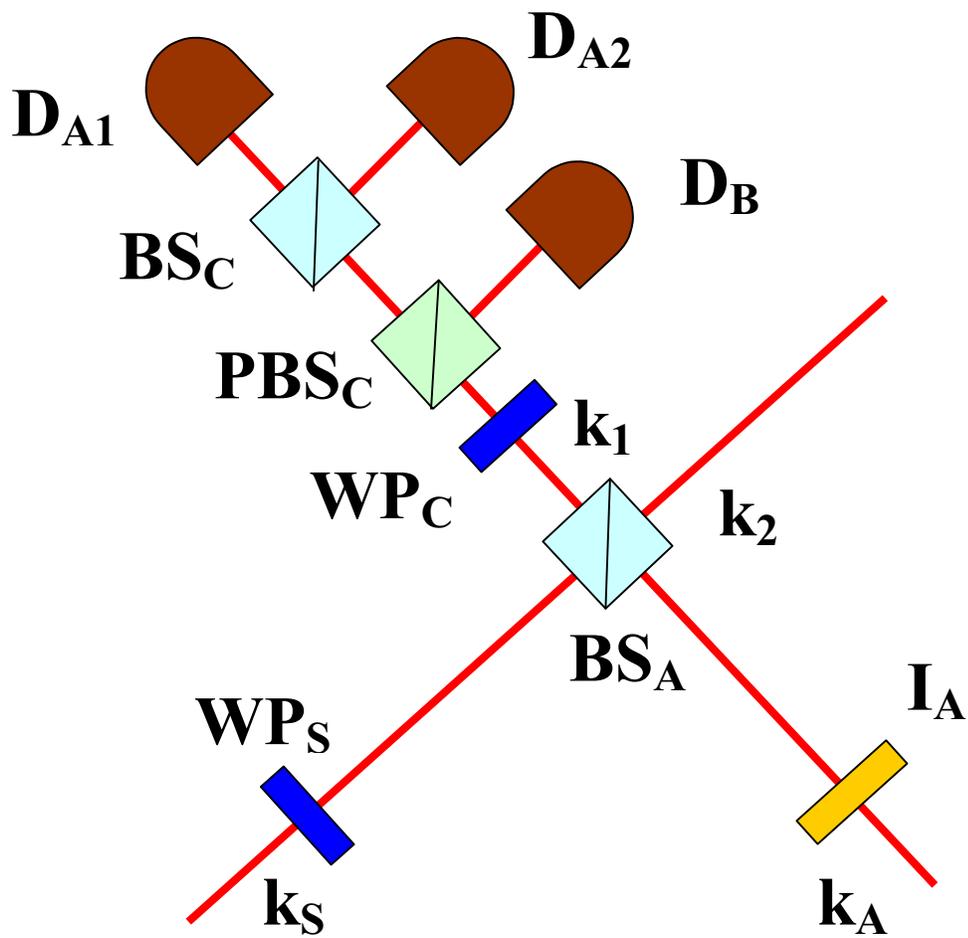
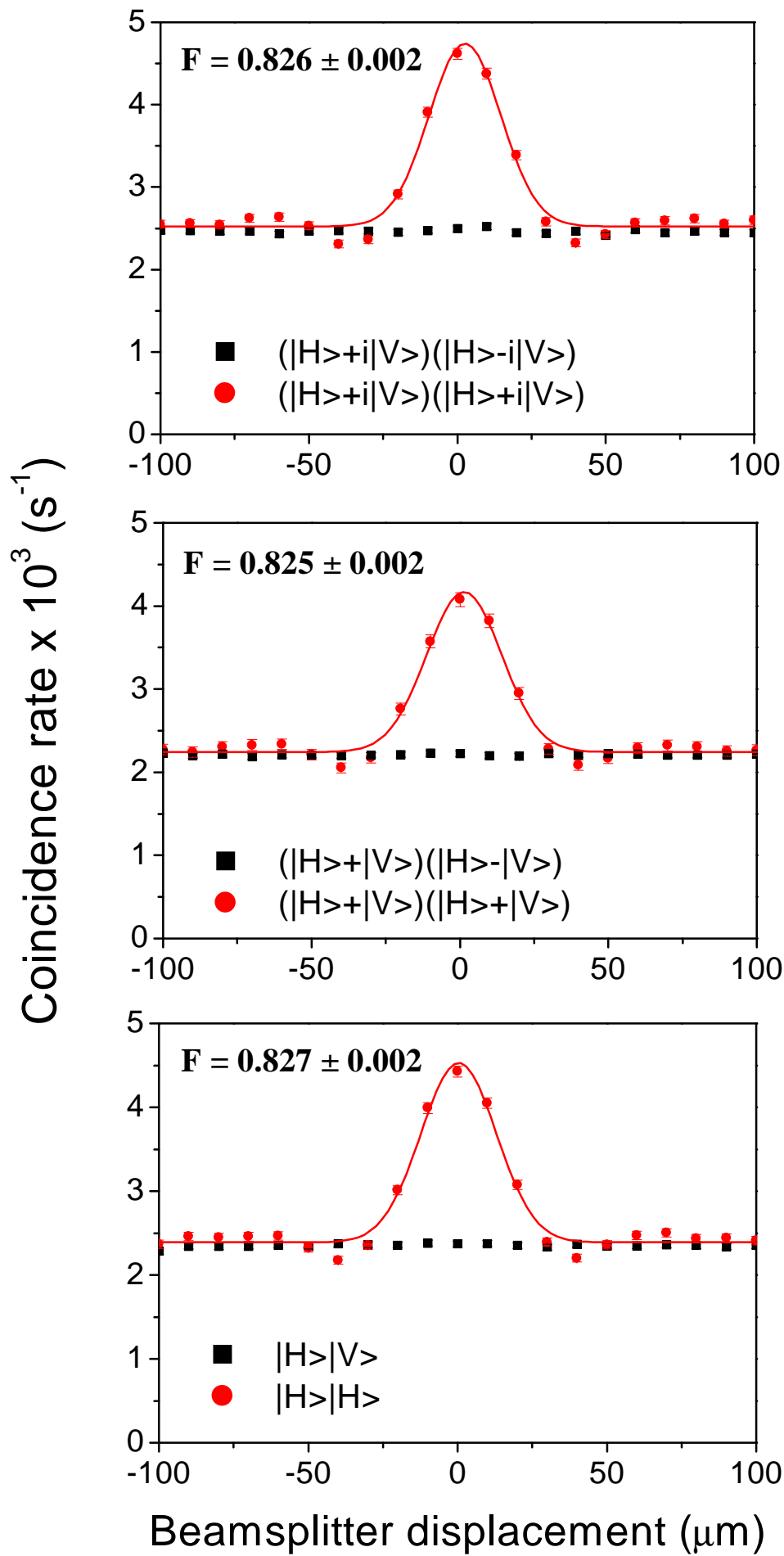
\end{document}